\documentclass[journal]{IEEEtran}
\usepackage{graphicx}
\usepackage{cite}
\usepackage{picinpar}
\usepackage{amsmath}
\usepackage{url}
\usepackage{flushend}
\usepackage[latin1]{inputenc}
\usepackage{colortbl}
\usepackage{soul}
\usepackage{multirow}
\usepackage{pifont}
\usepackage{color}
\usepackage{alltt}
\usepackage[hidelinks]{hyperref}
\usepackage{enumerate}
\usepackage{siunitx}
\usepackage{breakurl}
\usepackage{epstopdf}
\usepackage{pbox}
\usepackage{amsmath,amsfonts}
\usepackage{algorithmic}
\usepackage{algorithm}
\usepackage{array}
\usepackage[caption=false,font=normalsize,labelfont=sf,textfont=sf]{subfig}
\usepackage{textcomp}
\usepackage{stfloats}
\usepackage{url}
\usepackage{verbatim}
\usepackage{graphicx}
\usepackage{cite}
\usepackage{romannum}
\usepackage[dvipsnames]{xcolor}
\usepackage{booktabs}
\usepackage{multicol}
\usepackage{multirow}
\usepackage{hyperref}
\usepackage{setspace}

\usepackage[caption=false]{subfig}
\newcommand{\oracle}[1][$\cdot$]{\texttt{oracle}}
\newcommand{\mixture}[1][$\cdot$]{\texttt{mixture}}
\newcommand{\proposed}[1][$\cdot$]{\texttt{separated}}

\newcommand{\informed}{\textit{informed X-UMX}}

\begin{document}

\title{Activity-Guided Industrial Anomalous Sound Detection against Interferences}

\author{{Yunjoo Lee}$^{*}$, {Jaechang Kim}$^{*}$, and Jungseul Ok \\
Pohang University of Science and Technology (POSTECH)
}

\maketitle

\begin{abstract}
We address a practical scenario of anomaly detection for industrial sound data, where the sound of a target machine is corrupted by background noise and interference from neighboring machines. 
Overcoming this challenge is difficult since the interference is often virtually indistinguishable from the target machine without additional information. 
To address the issue, we propose SSAD, a framework of source separation (SS) followed by anomaly detection (AD), which leverages machine activity information, often readily available in practical settings. 
SSAD consists of two components: (i) activity-informed SS, enabling effective source separation even given interference with similar timbre, and (ii) two-step masking, robustifying anomaly detection by emphasizing anomalies aligned with the machine activity. 
Our experiments demonstrate that SSAD achieves comparable accuracy to a baseline with full access to clean signals, while SSAD is provided only a corrupted signal and activity information. 
In addition, thanks to the activity-informed SS and AD with the two-step masking, SSAD outperforms standard approaches, particularly in cases with interference. 
It highlights the practical efficacy of SSAD in addressing the complexities of anomaly detection in industrial sound data.
\end{abstract}


\definecolor{limegreen}{rgb}{0.2, 0.8, 0.2}
\definecolor{forestgreen}{rgb}{0.13, 0.55, 0.13}
\definecolor{greenhtml}{rgb}{0.0, 0.5, 0.0}

\section{Introduction}
\begin{figure}[!t]
\centering
\includegraphics[width=0.8\linewidth]{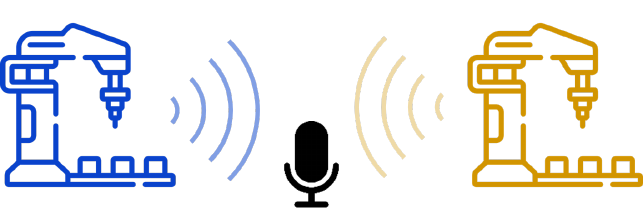}
\caption{The target situation where the machine sounds with similar acoustic features are recorded by a single-channel recording device.}
\label{fig:problem_illustration}
\vspace{-2mm}
\end{figure}

Anomalous sound detection is identifying irregularities that significantly deviate from the normal sound data.
Anomalous sound detection is widely applied, including industrial machinery inspection~\cite{ad_dereverberation, ad_smart_factory}, traffic monitoring~\cite{li2018anomalous}, and surveillance systems~\cite{crocco2016audio}, due to the cost-effectiveness and extensive coverage of auditory sensors.
In particular, within manufacturing plants, it holds great potential.
Audio sensors can detect malfunctioning components even from outside the machinery, aiding in accident prevention
However, industrial anomalous sound detection presents significant challenges, especially due to the presence of noise and interference that have similar acoustic characteristics, \emph{e.g.}, in the cases where the same type of machines operate simultaneously within a factory. 
Fig.~\ref{fig:problem_illustration} illustrates the scenario we targets where a single-channel recording device captures both the target sound and interference of similar acoustic features simultaneously.

To tackle this challenge, we propose a framework that combines Source Separation (SS) followed by Anomaly Detection (AD), referred to as SSAD.
This approach leverages activity signals to enhance source separation, allowing for effective isolation of the target machine's sound amidst similar interfering noises. 
The separated target machine is then passed on to the following anomaly detection module for analysis.  
Therefore, anomaly detection is less interrupted by the interference.
Also, we additionally utilize activity information in the anomaly detection process. 
Our anomaly detection model operates on the masked separated source, which is masked by a binary activity signal known as ``first-step masking".
After processing the input, the model calculates a weighted anomaly score, emphasizing active segments over inactive ones referred to as ``second-step masking". 
This process constitutes what we refer to as ``two-step masking."

Our main contributions are summarized as follows: 
\begin{itemize}

\item We integrate source separation into anomaly detection to tackle interference in industrial anomalous sound detection, thereby creating SSAD framework.

\item Our experiments illustrate the effectiveness of incorporating activity information,  in isolating interference when dealing with sources that share similar acoustic characteristics.

\item We stand out by introducing and applying the two-step masking, effectively leveraging activity information to enhance anomaly detection performance.

\item
Building upon our previous work~\cite{kim2023activity}, we extend our research with additional experiments to verify the robustness of source separation (Section~\ref{subsec:robusteness_ss}) and  of anomaly detection when dealing with additional sources (Section~\ref{subsec:AD_more_source}).

\item
We further analyze the effectiveness of the "two-step masking" technique (Sections~\ref{subsec:Two-Step1} to~\ref{subsec:Two-Step2}), and include comprehensive ablation studies (Section~\ref{subsec:ablation_studies}).

\end{itemize}

\IEEEpubidadjcol

\section{Related Work}
\subsection{Anomalous Sound Detection}
Sound-based anomaly detection is widely studied for its practicality and efficiency~\cite{idnn, guan2023anomalous}. 
Like common analysis methods for sound data, spectrogram is widely used for anomalous sound detection~\cite{bayram2021real, duman2020acoustic}.
In machine-learning based approaches of anomaly detection, using auto-encoders is a prominent approach~\cite{marchi2015novel, tagawa2015structured, idnn}.
When an auto-encoder model is trained on normal data, reconstruction errors of normal data and abnormal data are different.
This approach is especially useful in environments lacking labeled data, offering a practical solution for real-world anomaly detection~\cite{marchi2015novel, tagawa2015structured, kawachi2018complementary, idnn}.

In industrial environments, detecting anomalous sounds is critical for machine condition monitoring. 
The complex industrial environment often introduces additional noise and interference from nearby machinery, demanding robust solutions that can distinguish between normal operational sounds and potential faults.
To address this challenge, blind dereverberation algorithms are used to enhance detection accuracy by preprocessing and removing environmental noise~\cite{ad_dereverberation}. 
Concurrent with our work, source separation techniques are being explored to filter out irrelevant noises before anomaly detection~\cite{shimonishi2023anomalous}.
This approach uses two types of source separation models: one for general machine categories and another for specific machines, detecting anomalies by identifying sound discrepancies under abnormal conditions. 
However, \cite{shimonishi2023anomalous} requires a growing number of models as the number of sources increases. 
In contrast, our model uses a single, scalable separation model that efficiently handles multiple mixed sources and easily adapts to increasing numbers. 
Additionally, similar to our approach, \cite{nishida2022anomalous} exploits machine activity information for the diagnosis of a single machine, by adopting activity estimation as an auxiliary task. 
Unlike this study, which only considers environmental noise in factory settings and single-machine diagnosis, our approach supports simultaneous recordings from multiple machines, enhancing its practical applicability in industrial environments.

\subsection{Informed Source Separation}
Recent deep learning approaches achieved noticeable performance improvement in source separation~\cite{ryosuke21xumx, luo2022music, rouard2023hybrid}.
Informed source separation utilizes additional side information related to the sources to improve the separation quality.
This approach proves particularly advantageous when dealing with a limited training dataset~\cite{schulze2021informed} and when encountering high levels of separation difficulty ~\cite{takahashi2022amicable}.
Commonly used types of side information are musical score~\cite{score17, ewert17score}, aligned lyrics~\cite{jeon20lyric}, activity~\cite{activity, hung2020multitask}, and video~\cite{video05, video18}.
For instance, \cite{hung2020multitask} utilizes source activity information by proposing a multitask structure of activity detection and source separation.
It is noteworthy that prior research primarily focuses on enhancing the separation of mixtures with distinct sources through the use of side information. 
However, our work demonstrates that even in challenging mixtures with indistinguishable sources, the utilization of side information proves effective in facilitating source separation.

\section{Method}
\label{sec:method}

SSAD is a framework that integrates Source Separation (SS) into an Anomaly Detection (AD) model.
This integration enables the separation of interference from the machine's sound of interest, facilitating the utilization of the separated clean signal for anomaly detection.
In this section, we first describe a source separation framework utilizing activity signals
(Section~\ref{subsec:source_separation} - Section~\ref{subsec:architecture_for_separation}),
followed by an examination of two methods where the activity mask is employed in anomaly detection (Section~\ref{subsec:masked_auto_encoder} and Section~\ref{subsec:masked_anomaly_score}).
Finally, we discuss how the separation model is combined with the anomaly detection model (Section~\ref{subsec:ss_ad}).

\begin{figure}[!t]
\centering
\includegraphics[width=\linewidth]{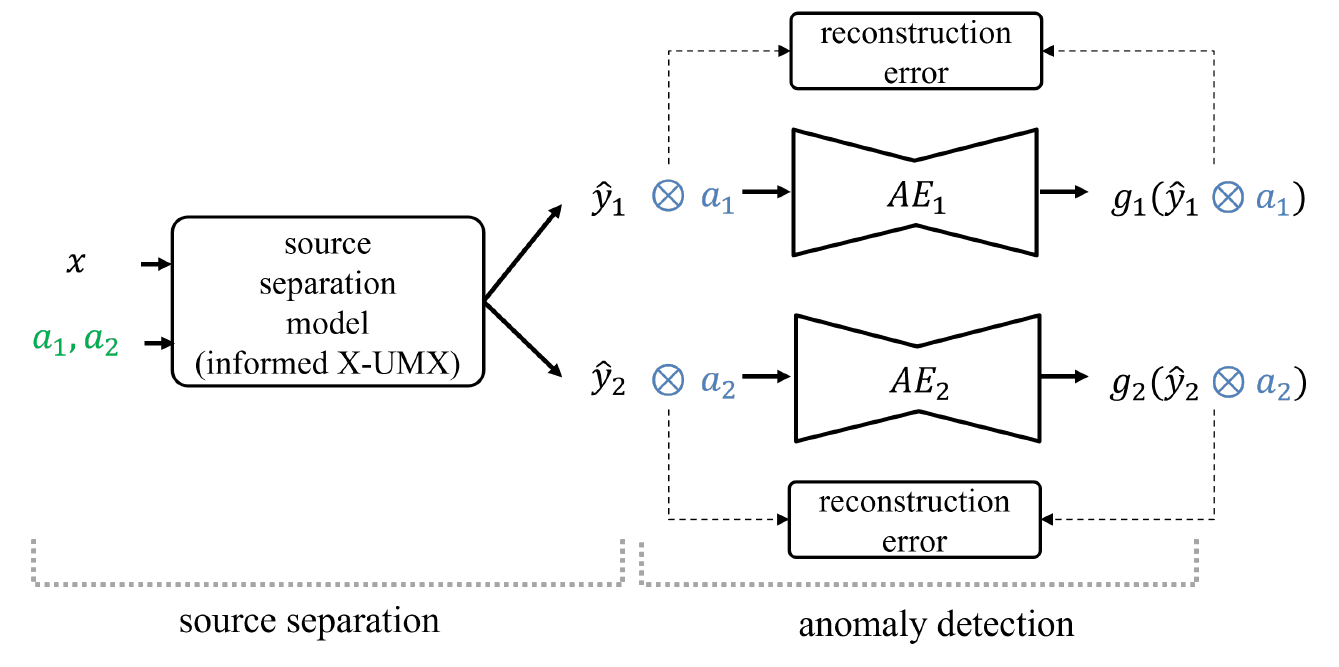}
\caption{An illustration of SSAD framework for anomaly detection of two machines, depicting the activity signals for source separation in {\color{ForestGreen} green} and the activity signals for two-step masking in {\color{Blue} blue}.}
\label{fig:SSAD}
\end{figure}

\subsection{Activity-informed Source Separation}
\label{subsec:source_separation}

Considering $k$ machines where each machine $i \in {1, ..., k}$ generates a source signal $y_i$, the mixture is denoted by $x$, defined as $x = \sum_{i =1}^k {y}_i$. 
Source separation then refers to the estimation of the individual sources $y_i$ from a mixture $x$ of $k$ number of sources.
To be specific, when $x$ is given to the source separation model $f = (f_1, ..., f_k)$, it separates $x$ into $\hat{y}_i$ such that, 
\footnote{A $\approx$ B denotes that A is an estimator of B. }
\begin{align}
   f_i(x) = \hat{y}_i \approx {y}_i.
\end{align}
Informed source separation, as referenced in~\cite{hung2020multitask, ewert17score, jeon20lyric}, incorporates additional data related to the sources during the separation process.
To be specific, we consider a simple scenario where binary activity information, indicating on/off state of the machine at each time step, is available.
Then the separation process with the assistance of the binary activity signal $a_i$ of machine $i$ is,
\begin{align}
\label{eqn:ss}
   f_i(x, a_i) = \hat{y}_i \approx {y}_i.
\end{align}
In addition, our focus is on single-channel source separation, utilizing a single-channel microphone for recording, which inherently presents greater challenges due to the absence of additional spatial information.

\begin{figure}[t]
    \begin{center}
    \includegraphics[width=1\linewidth, trim={0cm, 0cm, 0cm, 0cm},
        ]{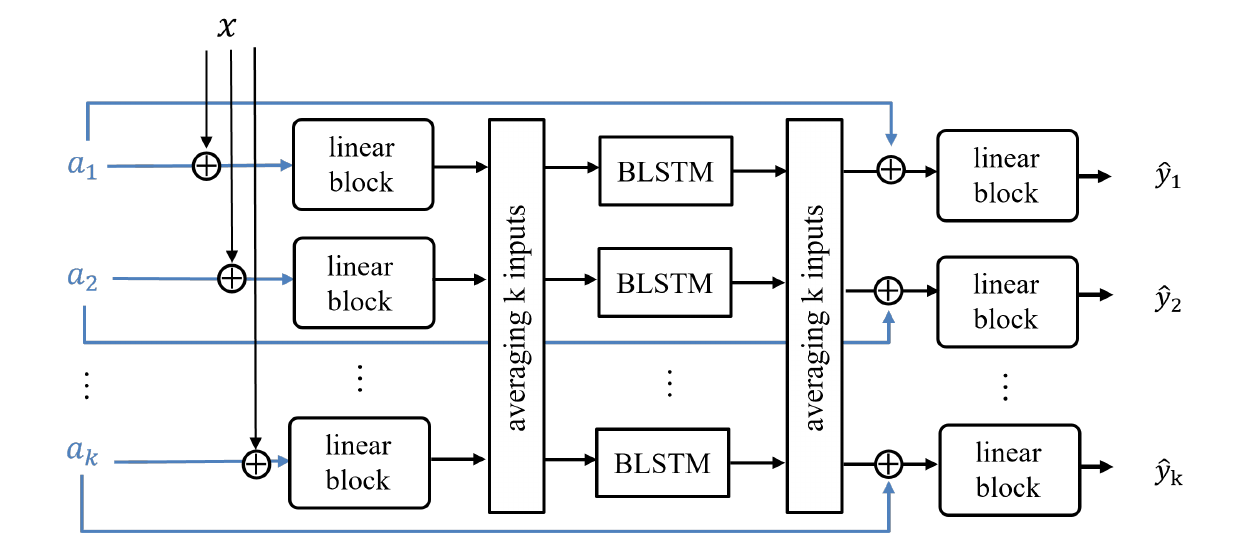}
    \end{center}
   \caption{Model architecture of the \informed, where the differences from the original X-UMX model is highlighted in {\color{Blue} blue}.  
   The model separates the $k$ number of sources based on the provided mixture $x$ and the activity signal $a_k$.}
   \label{fig:separation_model}
    \vspace{0.2cm}
\end{figure}

\subsection{Network Architecture for Activity-informed Source Separation}
\label{subsec:architecture_for_separation}    
We modify X-UMX\cite{ryosuke21xumx} for activity-informed source separation, which is originally designed for source separation without side information.
The overall architecture of modified model, termed  \emph{Informed X-UMX}, is illustrated in Fig.~\ref{fig:separation_model}, and the differences from the original X-UMX model are highlighted in blue.
The original X-UMX is a masking-based separation model, generating estimated masks $M_i$ for each source $i$. 
The masks are multiplied with the input mixture spectrogram  $|X| = \mathrm{STFT}(x)$ 
to obtain separated siganl $|\hat{Y_i}| = \hat{M_i} \otimes |X|$.
In the \emph{Informed X-UMX}, during the separation of the signal $y_i$ from the mixture $x$, we improve source separation by channel-wise concatenating the activity signal $a_i$ with the mixture at both the beginning and penultimate layers.

\begin{figure}[!t]
\centering
\includegraphics[width=60mm]{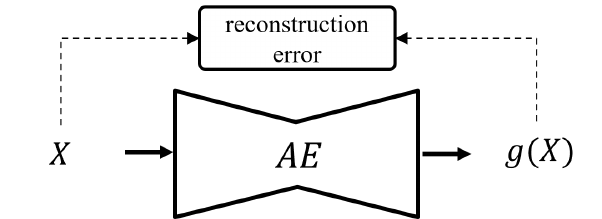}
\vspace{2mm}
\caption{The baseline anomaly detection model of a mixture using reconstruction error.}
\label{fig_1}
\end{figure}

\subsection{Activity Masked Auto-encoder for Anomaly Detection}
\label{subsec:masked_auto_encoder}
To detect anomalous sounds, we utilize a reconstruction-based anomaly detection technique as detailed in \cite{purohit19mimii}, employing auto-encoders (AE). 
These auto-encoders are trained on data representing normal states, leading them to reconstruct signals based on learned normal patterns. Consequently, when presented with abnormal input, significant discrepancies between the input and output signals may arise.
To be specific, each source $i$ has its own auto-encoder $g_i$, trained to reconstruct the input signal $\hat{y}_i$ such that,
\begin{align}
   g_i(\hat{y}_i) \approx \hat{y}_i.
\end{align}
For activity masked auto-encoders, we train individual auto-encoders $g_i$ for each source $i$ with the {\it masked} input signal  $a_i \otimes \hat{y}_i$ such that, \footnote{$\otimes$ is Hadamard product.}  
\begin{align}
   \tilde{y}_i \approx g_i(a_i \otimes \hat{y}_i) = g_i(\tilde{y}_i).
\end{align}
We refer to this masked auto-encoder as the ``first-step masking."
In addition, auto-encoders use mel-spectrograms as input, instead of a 1-dimensional signal, and
the activity mask has the same dimension as that of the input signal and then performs element-wise multiplication between them before being fed into the auto-encoders.

\subsection{Masked Anomaly Score}
\label{subsec:masked_anomaly_score}
Originally, the reconstruction error is used as a measure of of abnormality degree, often referred to as an anomaly score.
The reconstruction error when the source spectrogram $\hat{y_i}$ is given to the AE $g_i$ is as follows:
\begin{align} \label{eq:general_anomaly_score}
    \text{A}_i({x}) &= \| \hat{y_i} - g_i(\hat{y_i}) \|_2.
\end{align}
We also incorporate activity signals into anomaly score calculations. 
To be specific, we define a masked anomaly score with AE $g_i$ and an activity mask $a_i$:
 \begin{align} \label{eq:masked-score}
    \text{A}_i(x) = \| a_i \otimes (\hat{y}_i - g_i(\hat{y}_i)) \|_2  \;.
\end{align}
Inactive refers to a state where a machine is not expected to generate sound due to non-operation. However, the potential limited capacity of auto-encoders may result in reconstructed outputs with non-zero values in inactive areas, causing confusion when calculating anomaly scores. 
Therefore, exclusively focusing on the reconstruction error within the active area, indicated by the multiplication of the activity signal (in $\text{A}_i(x)$), helps remove unexpected noise from auto-encoders. We denote this masked anomaly score as ``second-step masking".

\subsection{Source Separation Followed by Anomaly Detection}
\label{subsec:ss_ad}
Our SSAD framework aims to  detect anomalies in each machine $i$ {\it individually} using a mixture signal $x = \sum_{i =1}^k y_i$, when the corresponding activity signal $a_i$ is also available.
To accomplish this, when the mixture $x = \sum_{i =1}^k y_i$ is given, we separate the source signal $y_i$ using the source separation model $f_i$ with the assistance of the binary activity signal $a_i$ of the machine $i$. This process yields $ f_i(x, a_i) = \hat{y}_i$ as in (\ref{eqn:ss}).
As shown in Fig.~\ref{fig:SSAD}, since each source $\hat{y}_i$ is separated from the mixture, we can diagnose each machine individually by training a set of auto-encoder $g_i$, which learns about $\hat{y}_i$ in normal state.
To incorporate activity information for anomaly detection, we utilize activity-masked auto-encoders, training them on $g_i(a_i \otimes \hat{y}_i)$.
Lastly, we adopt masked anomaly score (second-step masking) to the output of masked AE (first-step masking), constituting a two-step masking process of the activity signal.
This involves calculating the masked anomaly score $\text{A}_i(x)$ using the formula:
\begin{align}
\text{A}_i(x) = | a_i \otimes (\tilde{y}_i - g_i(\tilde{y}_i)) |_2,
\end{align}
where $\tilde{y}_i$ represents $a_i \otimes \hat{y}_i$.
The masked AE (first-step masking) is particularly effective for SSAD, as the separated signal might contain artifacts resulting from an imperfect source separation process.
By multiplying the activity signal before feeding it into auto-encoders, potential noise from the source separation is removed, reducing the volatility of judgments regarding the machine's condition.

\section{Experimental Settings}
\subsection{Dataset}
We use the 6dB signal-to-noise ratio split from the MIMII Dataset \cite{purohit19mimii}, which contains sounds from four types of industrial machines: fan, pump, slider, and valve. The MIMII dataset serves as the benchmark for sound-based machine anomaly detaction tasks.
To simulate real-world scenarios where similar machines operate simultaneously in factory settings, mixtures are generated by combining two sources of the same type.
For simplicity, we focus on slider and valve data, since they align with our assumption that the signal has both active and inactive areas, while fan and pump are always active. 
To increase the overlap ratio, we intentionally shift the sources to be more overlapped.
We define the overlap ratio as the ratio of simultaneously active sources to the active region of the mixture.
The overlap ratio for the mixture of two sources in the valve data is 0.2578, and for the slider, it is 0.3857.
Also, we synthesize binary activity signals based on the Root Mean Square (RMS) of a signal.
The threshold for activity labels of the signal is experimentally chosen considering the behavior of the target machine. 
For valve data, we assign 0 to the bottom  $20\%$ of the RMS values and 1 to the others, and for slider data, we use the mean value between the minimum and maximum of RMS as a threshold to assign active labels.
We create an anomalous mixture by incorporating one abnormal target source while maintaining the interference source as normal.
More details are available in our source code.

\subsection{Baseline Anomaly Detection Methods}
SSAD framework consists of two modules: source separation and anomaly detection. 
In our experiments, we validate and compare the configurations of each module. 
Depending on the input to the anomaly detection model, we consider a set of configurations in the following:
\vspace{-0.1cm}
\begin{itemize}
\setlength\itemsep{0.1cm}
    \item \oracle{} is a baseline using a clean source signal of the target machine as an input to train an auto-encoder $g_i$ and needs no source separation. 
    This would provide an upper-bound accuracy.
    The anomaly score for this case is  $\| {y_i} - g_i({y_i}) \|_2$.
    \item \mixture{} uses mixture as an input to train an auto-encoder $g_i$ for source $i$ and no source separation. 
    The diagnosis of the state of machine $i$ is made by exploiting activity information $a_i$,
    such that $\| a_i \otimes (x \otimes a_i - g_i(x\otimes a_i)) \|_2$.
    This is a baseline method most affected by the complexity of the mixture.
    \item \proposed{} is the proposed method using the output of the \informed{} to train an auto-encoder $g_i$.
     The masked anomaly score used is $\| a_i \otimes (\tilde{y}_i - g_i(\tilde{y}_i)) \|_2$.
\end{itemize}
However, we applied two-step masking for every configuration as a baseline and compared it with a no-masking model to validate the effectiveness of the masking.

\subsection{Evaluation Metrics}
As a performance metric of source separation, we use Signal-to-Distortion Ratio (SDR) defined as:
\begin{equation}
    \text{SDR}(s, \hat s) = 10 \log_{10} 
        \frac{{\| s \|} ^ 2}
        {{\| s - \hat s \|}^ 2} \;,
\end{equation}
where $s$ is the target signal and $\hat s$ is the predicted signal from mixture.
We compare anomaly detection methods using Area Under the Curve - Receiver Operating Characteristics (AUC-ROC, or shortened to AUC). 
AUC is the area under ROC curve varying detection threshold of anomaly score, which is plotted with $x$-axis of false positive rate and $y$-axis of true positive rate.
We note that random guess has AUC $0.5$ and a perfect model implies AUC $1.0$.

\section{Results}

        \begin{table*}[t]
            \caption{
            The AUC values for variants of SSAD.
            The column of \emph{no masking} denotes the variants of SSAD, where
            no masking is used for AE training and anomaly score.
            The column of two-step maskings is the proposed configuration 
            with the two-step maskings.
            Results are averaged over $10$ trials, and the 90\% confidence interval is provided in parentheses.
            The best AUC achieved from mixture data is shown in {\bf boldface}.
            }
            \vspace{3mm}
            \centering
            \resizebox{0.5\linewidth}{!}{
                \begin{tabular}{cc||cc}
                    \toprule
                    Method & 
                    data 
                    & no masking
                    & two-step masking
                    \\
                    \midrule
                    \midrule
                    \multirow{2}{*}{\shortstack{\\\oracle}} 
                    & valve 
                    & 0.655 ($\pm$ 0.018)
                    & 0.887 ($\pm$ 0.025)
                    \\
                    \cmidrule{2-4}
                    & slider 
                    & 0.903 ($\pm$ 0.024)
                    & 0.932 ($\pm$ 0.025)
                    \\
                    \midrule
                    \multirow{2}{*}{\shortstack{\\\mixture}} 
                    & valve mixture
                    & 0.496 ($\pm$ 0.043)
                    & 0.724 ($\pm$ 0.055)
                    \\
                    \cmidrule{2-4}
                    & slider mixture
                    & 0.794 ($\pm$ 0.040)
                    & 0.854 ($\pm$ 0.040)
                    \\
                    \midrule
                    \multirow{2}{*}{\shortstack{\\ \proposed}} 
                    & valve  mixture
                    & 0.626 ($\pm$ 0.046)
                    & \textbf{0.751} ($\pm$ 0.054)
                    \\
                    \cmidrule{2-4}
                    & slider  mixture
                    & 0.804 ($\pm$ 0.052)
                    & \textbf{0.899} ($\pm$ 0.038)
                    \\
                    %
                    \bottomrule
                \end{tabular}
            }
            \label{tab:main_auc}
        \end{table*}
   
\subsection{Performance Evaluation of SSAD}
Table~\ref{tab:main_auc} shows the effectiveness of SSAD (i.e., \proposed{}) in our target problem to detect anomalies with noise of similar timbre.
The first observation is that the \mixture{} suffers from a performance drop when compared to \oracle{}.
\mixture{} simulates a real-world noisy scenario in which the sounds of two machines of the same type are recorded simultaneously as a mixture.
\oracle{} represents the upper bound of the current anomaly detection model performance, as it is trained using clean source data.
The low AUC obtained from \mixture{} implies that interference from other machines makes the signal more complex and interrupts the assessments of the abnormalities.
Meanwhile, in \proposed{}, source separation before anomaly detection removes interference, making detection easier since the separated signal is less noisy and resembles clean signal estimates.
\begin{figure}[!t]
\centering
\includegraphics[width=\linewidth, trim={0cm, 0cm, 0cm, 0cm},]{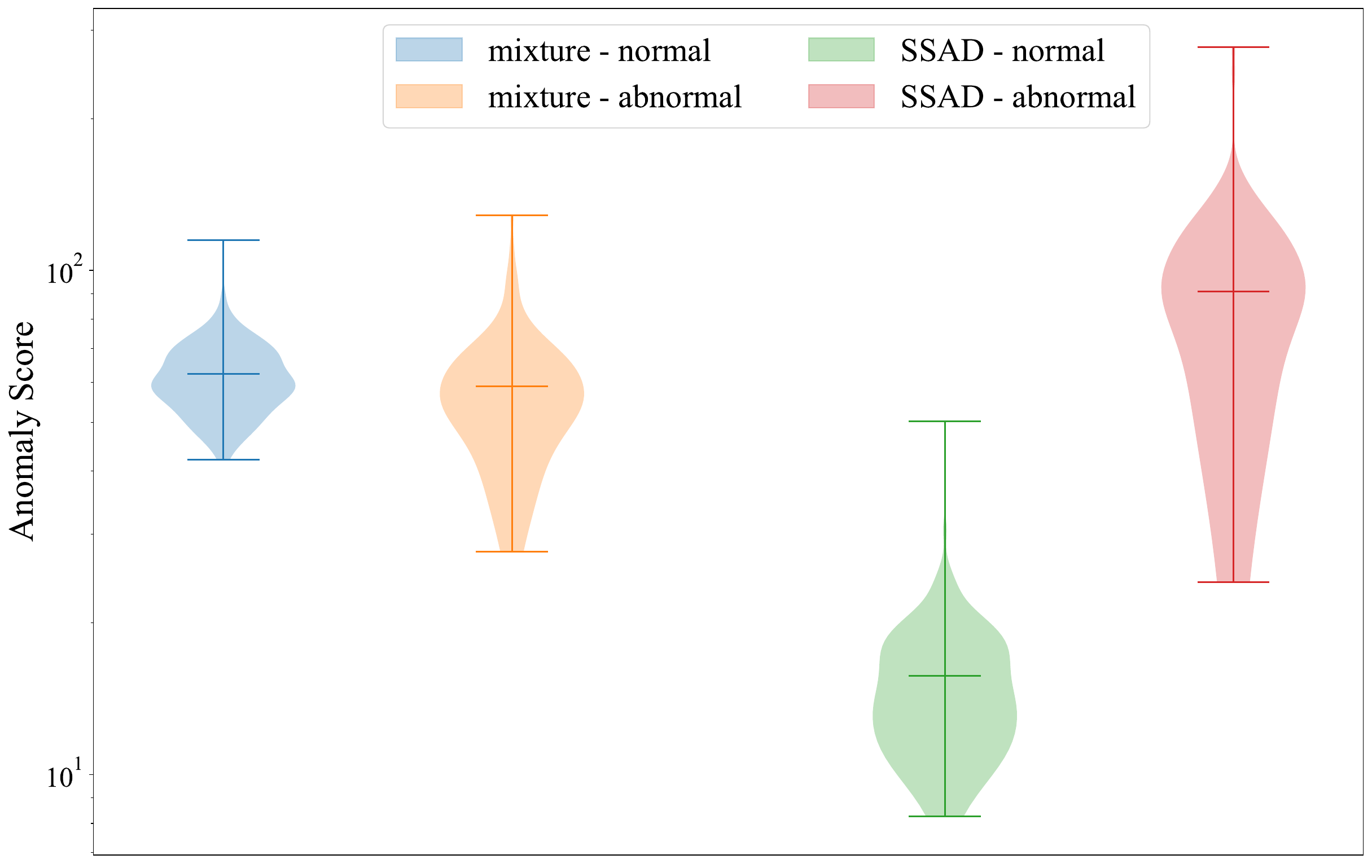}
\caption{Anomaly score distribution of valve data in normal and abnormal data.
Minimum, mean, and maximum values are marked.}
\label{fig:score_dist}
\end{figure}
We further explore why source separation improves anomaly detection by visualizing the anomaly score distribution in Fig.~\ref{fig:score_dist}.
For \mixture{} baseline, 
the anomaly score distributions of normal and abnormal datasets seem very similar and difficult to distinguish with a threshold. 
On the other hand, the distribution of \proposed{} for abnormal data noticeably differs from that of normal data, with a smaller overlap between them, facilitating easier threshold selection.

\subsection{Activity-informed Source Separation Performance}

\begin{table}[!t]
\caption{ 
    The source separation performances when the sources are from the same type of machines.
    }
    \label{tab:ss_model_comaprison}
\centering
\resizebox{\linewidth}{!}{
\begin{tabular}{c|c|c}
\hline
Model & valve SDR (dB) & slide rail SDR (dB) \\
\hline
original X-UMX & 1.7 & 1.6 \\
\hline
original X-UMX + PIT & 2.6 & 2.1 \\
\hline
{\textit{informed X-UMX}} & 8.3 & 6.2 \\
\hline
\end{tabular}}
\end{table}
In Table~\ref{tab:ss_model_comaprison}, we evaluate the performance gain by activity information in the source separation.
We test whether the {\textit{informed X-UMX}} could separate mixtures of the same type and compare its performance with that of the original X-UMX and PIT~\cite{yu17pit}.
The original X-UMX fails to separate sources with similar timbre, which shows SDR of 1.7dB and 1.6dB in valve and slider, respectively.
This limitation arises from the assumption of the original X-UMX, that the mixture comprises different types of sources with distinct timbre. 
When presented with a mixture of sources that have similar acoustic features, it struggles to separate the sources.
PIT~\cite{yu17pit} is a widely used approach for speech separation, where multiple speakers can be separated without using side information and explicit source index. 
Although PIT is known to be useful for multi-speaker separation, PIT only achieves an SDR gain of 0.9dB for the valve and 0.5dB for the slider. 
As SDR of the \emph{informed X-UMX} suggests, within the setting where timbres are difficult to distinguish, the activity information is effective in separating the sources. 

\label{subsec:further_analysis}
\begin{figure}[!t]
\centering
\includegraphics[width=1.0\linewidth, trim={1cm, 0cm, 0cm, 0cm}]{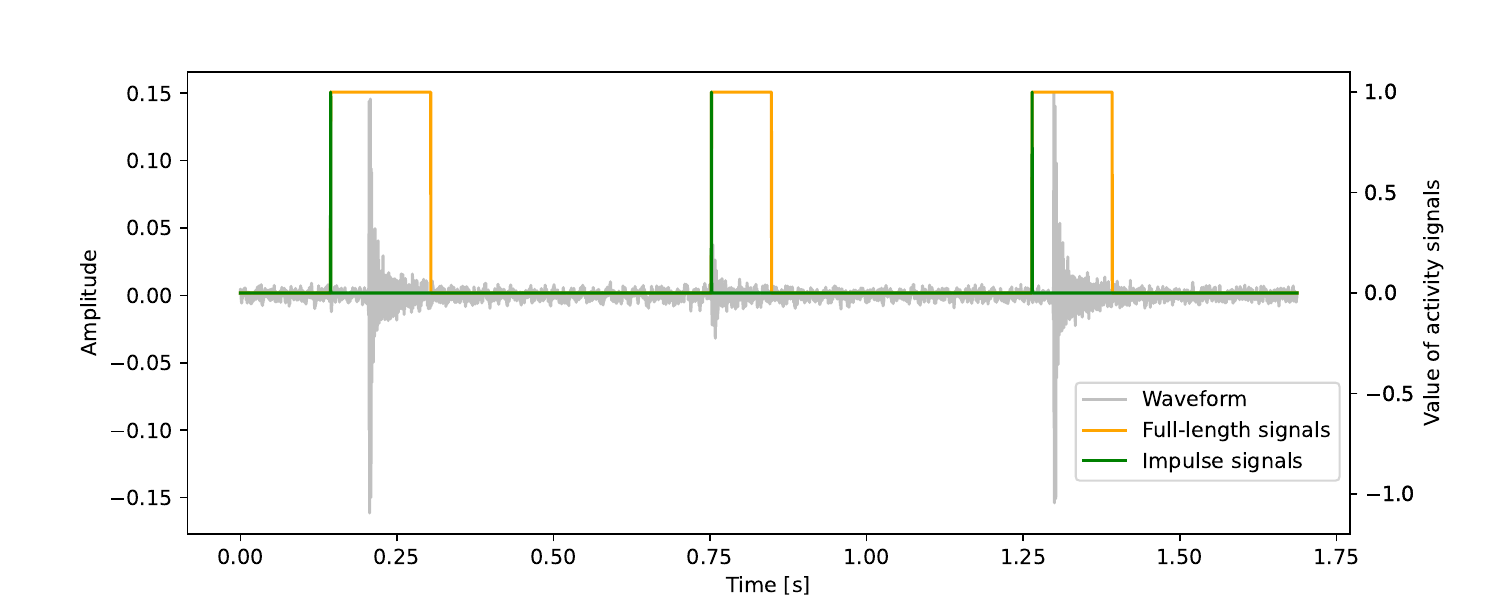}
\caption{The waveform of a valve (in {\color{darkgray} gray}) alongside its corresponding activity signals. These signals include the baseline full-length signal (depicted in {\color{Yellow} yellow}) and the simplest impulse signal (depicted in {\color{Green} green}). The y-axis displays the scale of the waveform on the left-hand side, while the scale of the activity signals, represented as a binary signal, is plotted on the right-hand side.}
\label{fig:impulse_signal}
\end{figure}

\subsection{Source Separation Robustness in Activity Signal Types}
\label{subsec:robusteness_ss}
This section explores the robustness of SSAD to incomplete activity signals, extending its applicability beyond binary activity.
We introduce incomplete activity signals named ``impulse signals", indicating only the start time of the machine without specifying when it stops, in contrast to ``full-length signals" which include both start and end times. 
Fig.~\ref{fig:impulse_signal} illustrates waveform, impulse signal, and full-length signal.
In our implementation, the signal is set to 1 upon machine startup and remains 0 until the next activation.
As shown in Table~\ref{tab:type_of_activity_signal}, source separation using impulse signals achieves significant improvement compared to scenarios without activity information.
Notably, the performance gap between impulse and full-length signals is relatively small. 
This suggests that the informed source separation model effectively utilizes activity signals even in a simplified form, demonstrating SSAD's applicability to various activity signal types.

\begin{table}[t]
\caption{Source separation performance analyzed across different signal activity durations.}\label{tab:type_of_activity_signal}
\centering
\resizebox{0.9\linewidth}{!}{
\begin{tabular}{c|c|c}
\hline
Activity signal type & valve SDR (dB) & slider SDR (dB) \\
\hline
no activity signal & 1.70 & 1.6 \\
\hline
impulse signal& {8.42} & {6.67} \\
\hline
full-length signal& {8.79} & {6.68} \\
\hline
\end{tabular}}
\vspace{-2mm}
\end{table}

\begin{figure*}[!t]
    \centering
    \subfloat[no masking auto-encoder]{%
    \includegraphics[width=0.27\textwidth, trim={0cm, -0.5cm, 0cm, 0cm}]{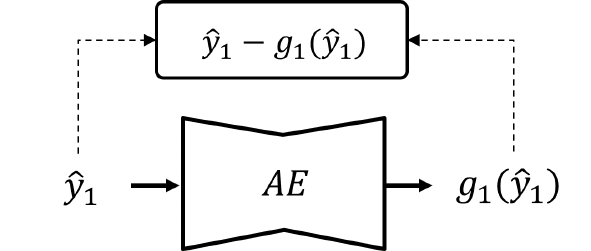}}
    \hfill
    \subfloat[one-step masking auto-encoder]{%
    \includegraphics[width=0.33\textwidth, trim={0cm, -0.5cm, 0cm, 0cm}]{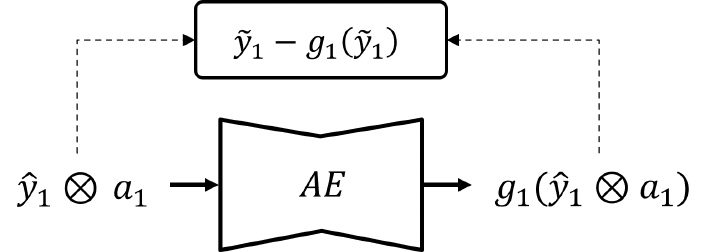}}
    \hfill
    \subfloat[two-step masking auto-encoder]{%
    \includegraphics[width=0.33\textwidth, trim={0cm, -0.5cm, 0cm, 0cm}]{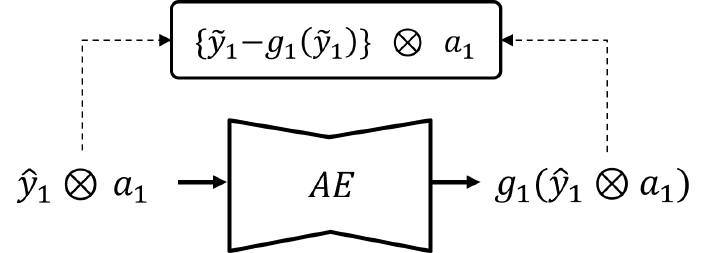}}
    \vspace{0.2cm}
    \caption{The variants of anomaly detection models with different masking strategies: no mask auto-encoder, one-step masking auto-encoder, and two-step masking auto-encoder.}
    \label{fig:ad_model}
\end{figure*}

\subsection{Anomaly Detection Robustness with More Sources}
\label{subsec:AD_more_source}
\begin{figure}[!t]
\begin{center}

\includegraphics[width=\linewidth, trim={0cm, 1cm, 0cm, 0cm},]{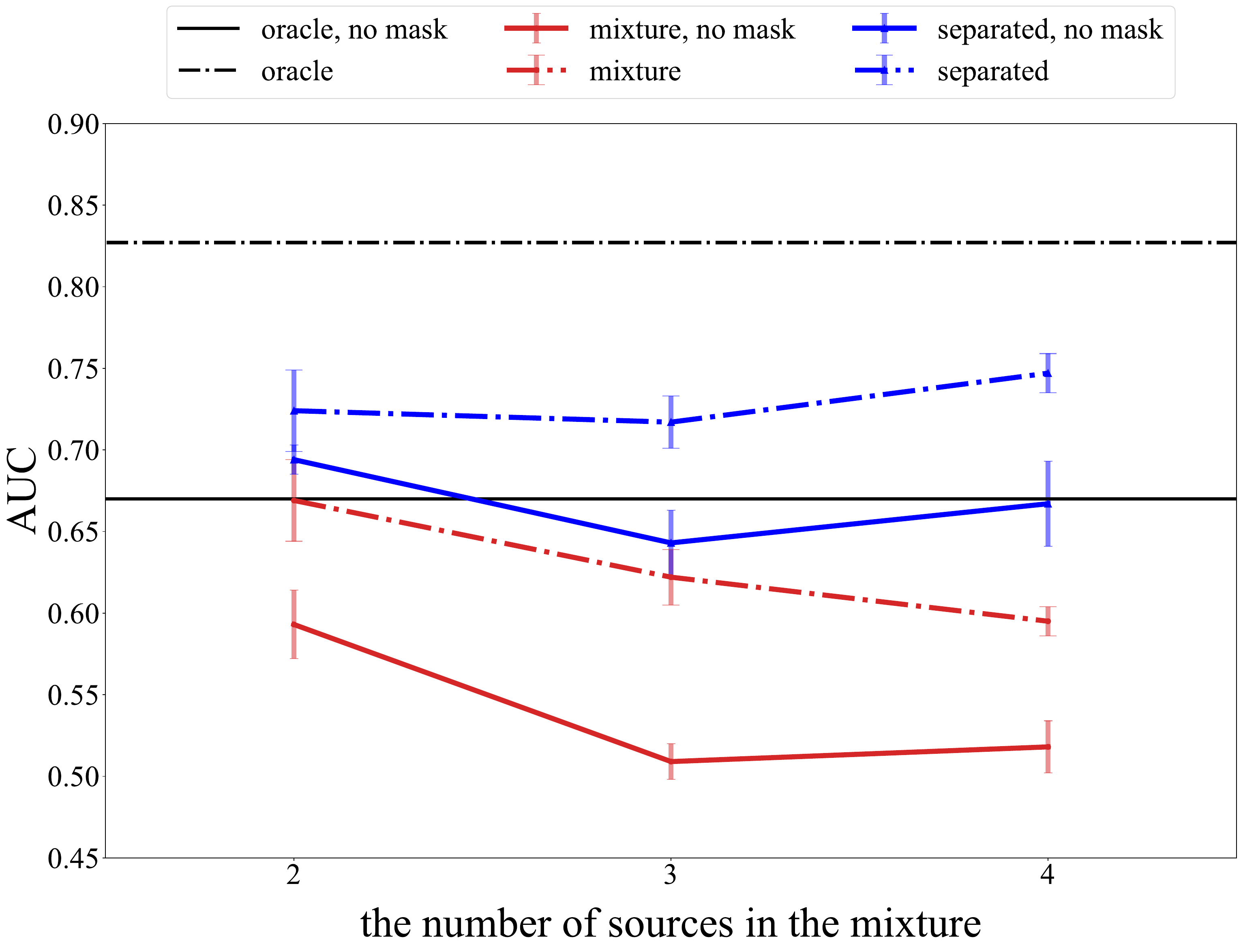}\label{fig:ad_more_sources}
\end{center}
\caption{
    An AUC comparison over the number of sources in the mixture. 
    Average SDR of the separation models are $8.3$, $7.0$, and $5.6$ dB.
    Error bars are 95\% confidence interval from 10 instances.
}
\label{fig:more_sources}
\end{figure}
Fig.~\ref{fig:more_sources} is the experimental results of more complicated mixture data. 
We additionally evaluate the mixture with three and four sources in valve data. When the number of sources increases the learning difficulty also increases, therefore we can observe the performance gap of \mixture{} and oracle along the number of sources increases. 
In contrast, the AUC of \proposed{} is more robust than \mixture{}.

\subsection{Two-Step Masking Effectiveness with Masking Variants}
\label{subsec:Two-Step1}
We evaluate the effectiveness of two-step masking against no masking and one-step masking, illustrated in Fig.~\ref{fig:ad_model}.
In the ``one-step masking" setting, the masked auto-encoder is used for anomaly detection, but masked anomaly score is not used.
In the ``two-step masking" setting, we additionally incorporate the masked anomaly score in the anomaly detection process, building upon the masked AE. 
In Table~\ref{tab:num_masks}, as the number of masking steps increases, the anomaly detection performance also improves.
For the valve data, when the two-step masking approach is utilized, the Area Under the Curve (AUC) is 0.751, which represents a significant improvement compared to the one-step masking setting, with a difference of 0.1.
Similarly, the slider dataset shows a 0.029 AUC increase with two-step masking.
These results suggest two key takeaways.
First, one-step masking effectively mitigates artifacts from potential source separation model imperfections, as evidenced by the performance improvement compared to no masking.
Second, the additional masking step in the two-step approach seems to enhance the model's ability to focus on critical data points, leading to a more efficient anomaly detection process. 
This is supported by the further performance boost observed when transitioning from one-step to two-step masking.

\begin{table}[!t]
\caption{The anomaly detection performance assessed using various masking variants.}\label{tab:num_masks}
\centering
\resizebox{0.9\linewidth}{!}{
\begin{tabular}{c|c|c|c}
\toprule
Source & no masking & one-step masking & two-step masking \\
\midrule
valve & 0.626 & 0.651 & 0.751\\
\midrule
slider & 0.783 & 0.870 & 0.899\\
\bottomrule
\end{tabular}}
\end{table}

\subsection{Two-Step Masking Effectiveness via Reconstruction Error}
\label{subsec:Two-Step2}
We compare the mean reconstruction error of the AE output at the active area and inactive area. 
A higher reconstruction error indicates that the auto-encoder fails to accurately reconstruct the input data, implying that it significantly deviates from the trained normal data. 
Consequently, anomalies are more likely to occur in that region.
The reconstruction error gaps between active and inactive states are 44.98 and 60.49 in the valve and slider datasets respectively.
This result suggests that by solely focusing on active regions through ``two-step masking", the model efficiently learns the inherent characteristics of abnormal machine behavior selectively.

\begin{table}[t]
   \caption{A comparison of reconstruction errors conducted between active and inactive states.}
    \centering
    \resizebox{0.45\linewidth}{!}{
        \begin{tabular}{c|c|c}
            \toprule
            machine & valve  & slider   \\
            \midrule
             active & 69.16 & 91.49 \\
            \midrule
            inactive & 24.18 & 31.00 \\
            \bottomrule
        \end{tabular}
    }

    \label{tab:table_active}
\end{table}


\vspace{-5pt}
\section{Ablation Studies}
\label{subsec:ablation_studies}
\subsection{Significance of AE Training Data}
In SSAD framework, the anomaly detection model is trained with the separated signal from the previous source separation module. 
Using ground truth single source data to train AE and evaluated with the separated source results in 0.589 and 0.775 AUC for separated valve and slider data respectively, both of which demonstrate lower performance compared to our proposed \proposed{}.
This supports our rationale that to judge whether the separated signal lies in the normal area, the decision should be made on the separated signal domain.
Lastly, it is noteworthy that to reduce misdetection from imperfect separation, the anomaly detection model needs to be trained with the separated signals (in normal status) rather than ground truth clean signals. 
Due to the possible imperfect source separation, the distribution of the separated signals in normal states is different from that of the clean signals. 
To judge whether the separated signal is statistically far from the trained normal signal, the decision should be made on the distribution of the separated signals not on the clean signal.

\begin{table}[!t]
\caption{An AUC comparison under abnormal interference considerations.}
\centering
\resizebox{0.75\linewidth}{!}{
\begin{tabular}{cc||cc}
\toprule
Method & input data & AUC\\
\midrule
\midrule

\multirow{2}{*}{\shortstack{\\\mixture}} 
& valve mixture
& 0.422
\\
\cmidrule{2-3}
& slider mixture
& 0.689
\\
\midrule         
\multirow{2}{*}{\shortstack{\\ \proposed}} 
& valve  mixture
& 0.821
\\
\cmidrule{2-3}
& slider  mixture
& 0.852
\\

\bottomrule
\end{tabular}
}
\vspace{-0.2cm}
\label{tab:second_anomaly}
\end{table}

\subsection{Robustness against Abnormal Interference}
In the previous experiments, we assume that the target source is either in a normal or abnormal state, while the interference is always in a normal state.
In this section, we consider the possibility of the interference being abnormal. 
When the target source is normal but the interference is abnormal, the anomaly detection model should indicate that the target source is in a normal state, irrespective of the interference state.
To be specific, we designed four scenarios: (normal, normal), (normal, abnormal), (abnormal, normal), and (abnormal, abnormal) for the target source and interference, respectively.
In an ideal scenario, the model should classify the first two cases as normal and the last two cases as abnormal.
We utilize two-step masked \mixture{} and \proposed{}  for comparison.
In Table~\ref{tab:second_anomaly}, the AUC of \mixture{} is 0.422 and 0.689 for the valve and slider data respectively, while the AUC of \proposed{} is 0.821 and 0.852. 
This implies that for the \mixture{}, when the target source is normal but the interference is abnormal, the area masked by the activity masks also includes the abnormal machine sounds.
As a result, the expected outcome of the target machine being normal is inaccurate.
However, for the \proposed{} method, when the interference is abnormal, the source separation module effectively separates the abnormal machine sounds from the interference source, making anomaly detection less challenging.
The SDR in this experiment setting is 7.15 and 5.13 for the valve and the slider datasets.

\section{Conclusion}
We address a challenging yet practical scenario of anomaly detection in the industrial environment, where machine sounds with similar acoustic features interfere with each other. 
To overcome these challenges, we propose SSAD, which leverages machine activity information through (i) informed source separation, and (ii) anomaly detection with two-step masking. 
Our experiments demonstrate that the proposed method, when provided with mixed signals and activity data, effectively separates the sources and makes judgments based on the separated results, achieving an accuracy comparable to that of the \oracle{} method trained exclusively on clean signals.
Also, we validated the effectiveness of two-step masking in anomaly detection through comprehensive ablation studies. 
Despite the performance improvement in anomaly detection achieved by our proposed method, further research is possible to address some limitations.
First, as our contributions primarily stem from the integration of source separation and anomaly detection, rather than the invention of a new architecture that optimally suits both, there remains room for improvement by enhancing both the source separation and anomaly detection models.
Furthermore, during testing, while we can diagnose the machine condition using only the mixture and activity signals of each machine, we still require the ground truth source signals for training the source separation model.
Addressing these limitations will be a crucial focus for our future work.


\bibliographystyle{ieeetr}
\bibliography{refs}

\end{document}